\documentclass[conference]{IEEEtran}
\IEEEoverridecommandlockouts

\usepackage{amsmath,amssymb}
\usepackage{breqn}
\usepackage{bm}
\usepackage[ruled,linesnumbered]{algorithm2e}

\usepackage{graphicx}
\usepackage{float}
\usepackage{subfigure}
\usepackage{multirow}
\usepackage{cite}

\def\BibTeX{{\rm B\kern-.05em{\sc i\kern-.025em b}\kern-.08em
    T\kern-.1667em\lower.7ex\hbox{E}\kern-.125emX}}

\begin{document}

\title{{RMTransformer: Accurate Radio Map Construction and Coverage Prediction}}

\author{
    \IEEEauthorblockN{Yuxuan Li\IEEEauthorrefmark{1}\IEEEauthorrefmark{2},
                     Cheng Zhang\IEEEauthorrefmark{1}\IEEEauthorrefmark{2},
                     Wen Wang\IEEEauthorrefmark{1}\IEEEauthorrefmark{2},  and
                      Yongming Huang\IEEEauthorrefmark{1}\IEEEauthorrefmark{2}}\\
\IEEEauthorblockA{\IEEEauthorrefmark{1} National Mobile Communications Research Laboratory, Southeast University, Nanjing, 210096, China \\ 
\IEEEauthorrefmark{2} Purple Mountain Laboratories, Nanjing, 211111, China \\
Email: yuxuan\_li@seu.edu.cn, zhangcheng\_seu@seu.edu.cn, wangwen@pmlabs.com.cn, huangym@seu.edu.cn}
}

\maketitle

\begin{abstract}
    Radio map, or pathloss map prediction, is a crucial method for wireless network modeling and management. By leveraging deep learning to construct pathloss patterns from geographical maps, an accurate digital replica of the transmission environment could be established with less computational overhead and lower prediction error compared to traditional model-driven techniques. While existing state-of-the-art (SOTA) methods predominantly rely on convolutional architectures, this paper introduces a hybrid transformer-convolution model, termed \textit{RMTransformer}, to enhance the accuracy of radio map prediction. The proposed model features a multi-scale transformer-based encoder for efficient feature extraction and a convolution-based decoder for precise pixel-level image reconstruction. Simulation results demonstrate that the proposed scheme significantly improves prediction accuracy, and over a 30\% reduction in root mean square error (RMSE) is achieved compared to typical SOTA approaches.

\end{abstract}

\section{Introduction}
Large scale fading, i.e., pathloss, characterizes the reduction in signal strength from a transmitter, and reflects the transmission environment and channel conditions of a wireless network. As a potential enabler for network management in 6G systems, digital twin (DT) aims to provide a digital replica of the real-world network.
This replica serves as a pre-validation and safe-exploration environment for various applications \cite{you_ai}, including network planning \cite{10121572, 9863238}, beamforming \cite{dtbf,conf,xiong}, localization \cite{noauthor_locunet_nodate}, resource allocation \cite{zhang_dt}, and scheduling \cite{9053347}. Accurate radio maps, in turn, form the foundational channel state information (CSI) necessary for DT modeling.

Traditional model-driven approaches, such as ray tracing and real-world channel measurements, face significant challenges in practical DT applications due to their high overhead and computational costs. In recent years, deep learning has emerged as a promising tool for radio map prediction, and significant improvements in prediction accuracy are achieved. However, most state-of-the-art (SOTA) methods rely heavily on convolutional neural networks (CNNs), while advanced models and architectures, such as transformers and vision attention mechanisms, remain underexplored. To address this gap, this paper introduces the latest multi-scale transformers to radio map prediction, and propose a hybrid transformer-convolution model, termed \textit{RMTransformer}, i.e., radio map transformer, to achieve highly accurate radio map prediction.

\subsection{Related Works}
\vspace{-0.18em}
To efficiently illustrate the wireless propagation environment, extensive efforts have been dedicated to radio map and DT construction, including both model-driven and data-driven approaches. A widely-used model for large-scale channels is the dominant path model \cite{dominant}. While these models rely on simplified assumptions, their performance is often inadequate in complex environments.
Another prevalent model-driven approach is ray tracing \cite{10198573, 9773088}, which has been demonstrated effective in simulation accuracy. However, the substantial computational complexity may limit its efficiency, especially considering the time constraints and the more intricate channel models anticipated in 6G, such as those involving high frequencies and large-scale network deployments.

To overcome the drawbacks of model-driven approaches, deep learning has been widely applied to wireless communications and channel modeling \cite{romero_radio_2022}. When the channel model is assumed to be available, a possible strategy is to utilize neural networks to simulate and represent the electromagnetic propagation.  For instance, the authors in \cite{WiNeRT} propose a neural surrogate to model wireless electromagnetic propagation effects in indoor environments. Similarly, the authors in \cite{zhao_nerf2_2023} utilize NeRF2 to construct a continuous volumetric scene function for interpreting RF signal propagation, while in \cite{wen2024wrfgswirelessradiationfield}, a model called WRF-GS is proposed for channel modeling based on wireless radiation field (WRF) reconstruction using 3-dimensional (3D) Gaussian splatting.  However, these assumptions on specific channel models are not necessarily satisfied in practical environments. When the channel model is biased and differs from the exact propagation environment, significant deteriorations in performance might occur.

For the radio map prediction problem, although reference models such as the 3GPP pathloss model \cite{3gpp_pathlpss} are available, the complex propagation environment in the practical networks may prevent us from obtaining accurate pathloss values solely from the models. The intricacies of real-world wireless environments make it nearly impossible to encapsulate all characteristics within a single simplified model. Therefore, a natural idea is to utilize fully data-driven approaches, impelling advanced neural networks to learn the complicated wireless propagations, e.g., to predict the radio map using geographical information through deep learning. Regarding to deep learning-based radio map prediction, typical SOTA approaches  include RadioUnet \cite{levie_radiounet_2021}, which employs an encoder-decoder structure based on CNNs for efficient feature extraction, and PMNet \cite{lee_pmnet_2023, lee_scalable_2024}, which further substitutes the conventional convolution blocks with Atrous convolution for additional improvement. According to the first competition for radio map prediction held in IEEE ICASSP 2023 \cite{yapar_first_2023}, PMNet achieves the highest prediction accuracy among the typical deep learning approaches.

\subsection{Contributions}
Generally, the currently dominant schemes for radio map prediction are still CNN-based supervised learning approaches. To explore alternatives beyond conventional CNNs, we investigate the potential of transformers and vision attention mechanisms to further enhance the prediction accuracy. Originally developed for natural language processing (NLP), transformers have been adapted for computer vision (CV) tasks, i.e., vision transformers (ViTs). ViTs have demonstrated superior performance in CV tasks such as image classification, demonstrating their excellent capability in feature representation. In this paper, we propose a transformer-enabled model for radio map construction, termed \textit{RMTransformer}, which integrates ViTs into the radio map prediction task. The primary contributions are as follows:
\begin{itemize}
    \item We design a hybrid transformer-CNN model, RMTransformer, based on an encoder-decoder structure. The transformer-based encoder efficiently extracts features from the geographical map, while the CNN-based decoder reconstructs the pixel-level radio map. Leveraging the powerful feature representation capabilities of ViTs, our approach significantly improves prediction accuracy compared to conventional CNN-based schemes. 
    \item We utilize a multi-scale transformer architecture in the encoder to generate features with different dimensions, so that the lower-dimensional representation of different scales could be learned for better feature extraction. These multi-dimensional features are then simultaneously propagated into the CNN decoder blocks with skip connection for image reconstruction to better accommodate the task of pixel-level image regression.
    \item We evaluate the proposed scheme on the public radio map dataset. Simulation results demonstrate that high prediction accuracy (e.g., $10^{-3}$ level of RMSE, and over 30\% improvement compared to PMNet) is achieved, indicating the desirable capability of the proposed scheme in delicately understanding the wireless environment.
\end{itemize}

\section{Problem Formulation}
Consider the geographical map $\mathbf{G}$ of an area served by a base station (BS) at location $m$. The environment layout information is described in $\mathbf{G}$, e.g., the regions of interest (RoIs) where user equipments (UEs) are probably deployed can be marked with $1$, while other areas like buildings or blockages are marked with $0$. For the UE at location $x \in \mathbf{G}$, the pathloss in dB is represented as 
\begin{equation}
    \mathrm{PL}(x, m) = {P_{\mathrm{RX}}(x, m)} - {P_{\mathrm{TX}}(m)},
\end{equation}
where $P_{\mathrm{TX}}(m)$ is the transmitted power from the BS, and $P_{\mathrm{RX}}(x, m)$ is the received power at the UE in dBm. The path loss is a fundamental description of the wireless propagation characteristics, which is related to factors including large-scale fading, small-scale fading, and shadowing, etc. The radio map $\mathbf{R}=\{P_{\mathrm{RX}}(x, m)\}_{x \in \mathbf{G}}$ represents the received power values at all locations within the geographical map, with $P_{\mathrm{RX}}(x, m)={P_{\mathrm{TX}}(m)} + \mathrm{PL}(x, m)$. For simplicity, the radio map is uniformly quantified into $H \times W$ square grids, each with a side length of $M$ meters. The pathloss value for a grid is represented as the average received power over the grid area. Therefore, the geographical map and radio map can be considered as 2D images of size $H \times W$, and a location $x$ in the map can be represented with a coordinate $(i, j)$, where $i=1, \dots, H$ and $j=1, \dots, W$.

There are several model-based approaches to describe the characteristics of pathloss, among which the most typical one is the 3GPP 38.901 channel model \cite{3gpp_pathlpss}, where the pathloss is expressed as
\begin{equation}
    \mathrm{PL}(x, m) = 10\alpha \log_{10}(d_{x,m}) + \beta + \mathrm{SF},
\end{equation}
where $d_{x,m}$ is the distance between the BS and location $x$, $\alpha$ and $\beta$ are coefficients related to the frequency and environmental characteristics, and $\mathrm{SF} \sim \mathcal{N}(0, \sigma^2)$ is the shadow fading factor. These parameters vary significantly depending on factors such as the geographical layout, building materials, blockages, and the presence of a line-of-sight (LoS) link. Such variability is difficult to capture using limited geographical information, and the models often only provide statistical approximations. In specific applications like network management or resource allocation, such statistical data may be insufficient, necessitating the construction of more accurate radio maps.

In this paper, we utilize neural networks to learn the mapping relationship between the geographical map $\mathbf{G}$ and radio map $\mathbf{R}$, i.e.,
\begin{equation}
    \hat{\mathbf{R}} = f_{\theta}(\mathbf{G}),
\end{equation}
where $\hat{\mathbf{R}}=\{\hat{P}_{\mathrm{RX}}(x, m)\}_{x \in \mathbf{G}}$ is the estimated radio map, $f_{\theta}$ is the mapping function learned from the neural network, and $\theta$ represents the neural network parameters. By designing appropriate neural networks and updating their parameters through the training dataset, we aim to minimize the error between the estimated radio map and the ground truth, i.e.,
\begin{equation}
    \min_{\theta} |\mathbf{R} - f_{\theta}(\mathbf{G})|.
\end{equation}

\begin{figure*}
    \centering
    \includegraphics[width=0.9\textwidth]{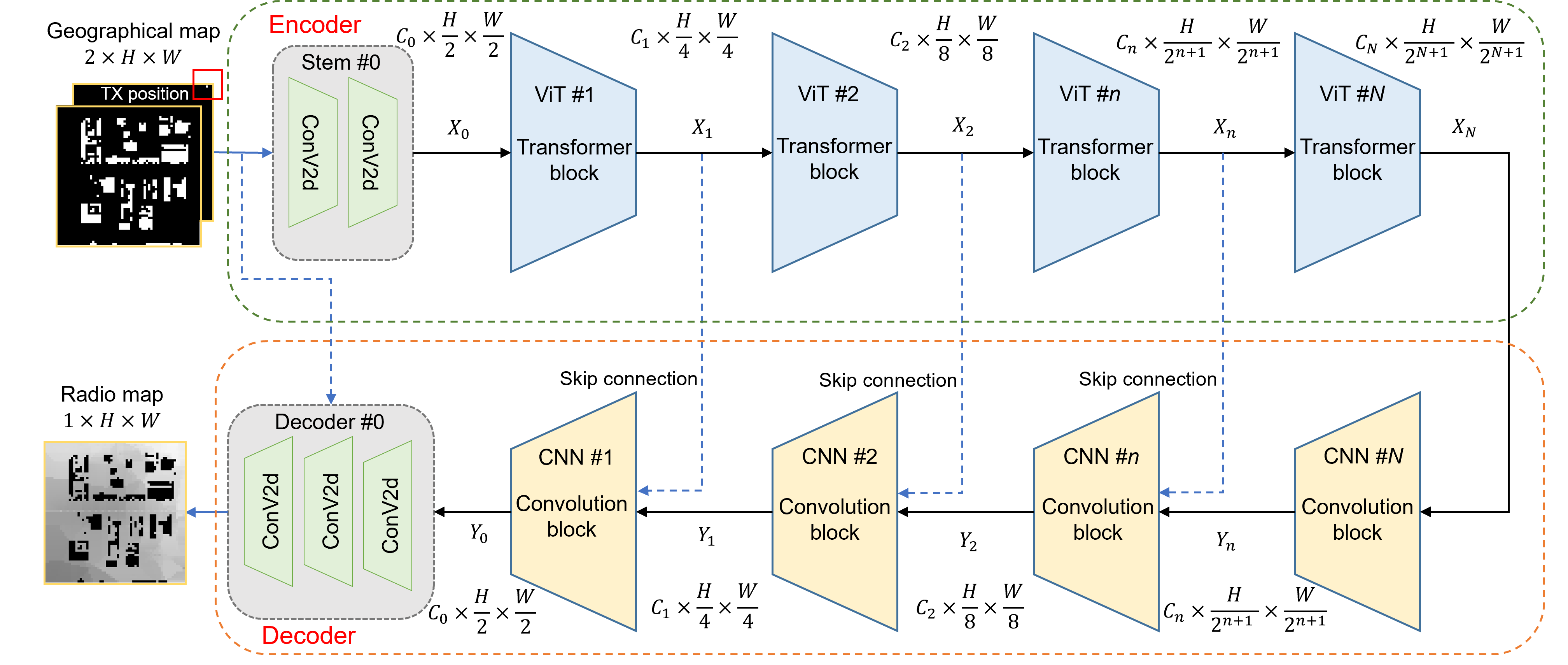}
    \caption{The architecture of RMTransformer.}
    \label{fig.radiotrans}
\end{figure*}

\section{Radio Map Prediction based on RMTransformer}
\subsection{Network Architecture}
To accurately construct the radio maps using the geographical information, we propose a transformer-based neural network architecture, called \textit{RMTransformer}. Similar to RadioUnet \cite{levie_radiounet_2021} and PMNet \cite{lee_scalable_2024}, we adopt an encoder-decoder structure, which facilitates learning low-dimensional representations of different scales from high-dimensional data. Unlike prior works that rely on CNNs as the backbone for both the encoder and decoder, we integrate the latest transformer and vision attention architectures into the neural networks for radio map prediction. Specifically, we propose a hybrid transformer-CNN model, where transformer blocks are employed in the encoder for feature extraction, while CNNs are utilized in the decoder for pixel-level image reconstruction. The overall architecture of RMTransformer is illustrated in Fig. \ref{fig.radiotrans}, which is composed of a CNN-based stem module to process the input image, $N$ transformer blocks in the encoder, $N$ CNN blocks in the decoder for multi-scale feature extraction, and an output CNN module for radio map image reconstruction. The details of these components are elaborated in the following subsections.

\subsection{Multi-Scale Transformers}
Transformers, powered by their self-attention mechanism, have achieved significant improvements in CV tasks, demonstrating their effectiveness in feature extraction and representation. Single-scale features are usually utilized in typical image classification tasks, e.g., the output of the transformer block is mapped to a vector indicating the corresponding classes using a multi-layer perceptron (MLP) header. However, pixel-level image regression tasks, such as radio map prediction, usually demand multi-scale feature representations for high precision. To meet this requirement, we extend the conventional transformer design into a multi-scale feature extractor.

In the encoder of RMTransformer, we utilize a series connection of $N$ transformer blocks for the multi-stage feature extraction with $N$ feature maps of different dimensions. Each transformer block $n, n=1, \dots, N$, takes the output feature map $X_{n-1} \in \mathbb{R}^{C_{n-1} \times H_{n-1} \times W_{n-1}}$ of the previous transformer block as its input, where $C_{n-1}, H_{n-1}$, and $W_{n-1}$ denote the number of channels, height, and width of the feature map, and generate the down-sampled feature map $X_{n} \in \mathbb{R}^{C_{n} \times H_{n} \times W_{n}}$, with $H_{n} = \frac{H_{n-1}}{2}$ and $W_{n} = \frac{W_{n-1}}{2}$. After the input data is processed through the $N$ transformer blocks, $N+1$ feature maps of different dimensions, i.e., dimensions of $\frac{H}{2} \times \frac{W}{2}, \frac{H}{4} \times \frac{W}{4}, \dots, \frac{H}{2^{N+1}} \times \frac{W}{2^{N+1}}$, could be derived, which is then propagated to the corresponding CNN blocks in the decoder with skip connection to realize accurate pixel-level radio map reconstruction.

\begin{table*}[hbt]
    \centering
    \caption{Neural network structure}
    \label{tab.net}
    \begin{tabular}{ccccccc}
        \hline
        \multicolumn{3}{c}{\textbf{Encoder}} & \multicolumn{4}{c}{\textbf{Decoder}} \\
        \hline
        \textbf{\#Module}     & \textbf{Description}              & \textbf{Feature dimension}         & \textbf{\#Module}     & \textbf{Description}     & \textbf{Feature dimension} & \textbf{Skip connection} \\
        Input  & Geographical map  & $2 \times 256 \times 256$ & Output & Radio map  & $1 \times 256 \times 256$ & -- \\
        0    & Conv2d  & $128 \times 128 \times 128$ & 0 & Conv2d  & $1 \times 256 \times 256$ & Encoder \#Input \\
        1    & MaxViT block  & $128 \times 64 \times 64$ & 1 & ConvTranspose2d  & $128 \times 128 \times 128$ & Encoder \#1 \\
        2    & MaxViT block  & $256 \times 32 \times 32$ & 2 & ConvTranspose2d  & $256 \times 64 \times 64$ & Encoder \#2 \\
        3    & MaxViT block  & $512 \times 16 \times 16$ & 3 & ConvTranspose2d  & $512 \times 32 \times 32$ & Encoder \#3 \\
        4    & MaxViT block  & $1024 \times 8 \times 8$ & 4 & ConvTranspose2d  & $1024 \times 16 \times 16$ & -- \\
        
        \hline
    \end{tabular}
\end{table*}
\begin{table}[hbt]
    \centering
    \caption{The prediction errors of different schemes}
    \begin{tabular}{cccc}
        \hline
        Scheme             & RadioUnet  & PMNet   & RMTransformer \\
        \hline
        RMSE               & 0.01904   &  0.01046 & \textbf{0.007148}  \\
        Ch. Pred. Err. & 0.02163   &  0.01186 & \textbf{0.008099}  \\
        Cov. Pred. Err.    & 0.0322  & 0.01774 & \textbf{0.01236}         \\
        \hline
    \end{tabular}
    \label{tab.nmse}
\end{table}

\subsection{Training Process}
Similar to \cite{levie_radiounet_2021} and \cite{lee_scalable_2024}, we take the 2-channel geographical map composed of the building and blockage map in the first channel and the location of BS in second channel as the input of RMTransformer. The input data is firstly processed by the CNNs in the stem module and a feature map $X_0 \in \mathbb{R}^{C_0 \times \frac{H}{2} \times \frac{W}{2}}$ is generated. Afterwards, $N$ different dimensional feature maps are derived through the $N$ transformer blocks, i.e.,
\begin{equation}
    X_{n} = g_{n} (X_{n-1}),\ \ n=1, \dots, N,
\end{equation}
where $g_{n}$ denotes the operations performed by the $n$-th transformer block. In this paper, we employ multi-axis vision transformer (MaxViT) blocks \cite{tu_maxvit_2022} in the encoder, which utilizes a block attention module and a grid attention module to balance the attention operations on both local and global data. Afterwards, multi-dimensional features could be effectively extracted, and each feature map $X_{n}$ is then propose into the $(n-1)$-th decoder CNN block in the decoder via skip connection together with the output of $n$-th decoder CNN block $Y_n$, i.e.,
\begin{equation}
    Y_{n-1} = h_{n-1} (\mathrm{concat}[X_{n}, Y_{n}]),\ \ n=1, \dots, N,
\end{equation}
where $h_{n-1}$ represents the operations of the $(n-1)$-th CNN block, and $\mathrm{concat}[X_{n}, Y_{n}]$ denotes the concatenation of $Y_{n}$ and $X_{n}$ in the skip connection. Finally, the pixel-level radio map image is reconstructed using the output CNN module.

\begin{figure*}[hbt]
    \centering
    \subfigure[RadioUnet]{
        \includegraphics[width=0.27\textwidth]{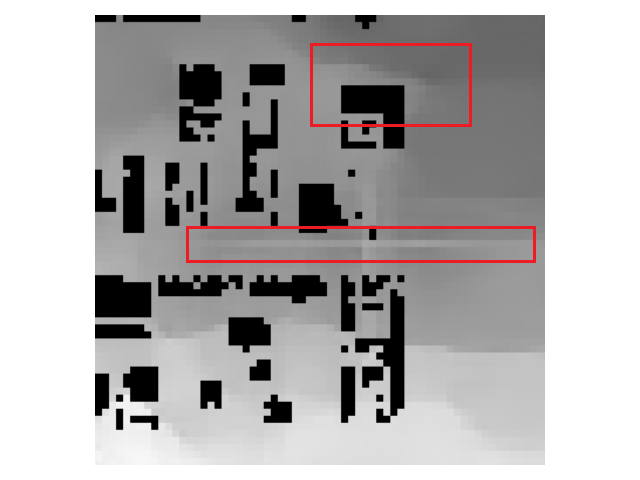}}
        \hspace{-3.18em} 
    \subfigure[PMNet]{
        \includegraphics[width=0.27\textwidth]{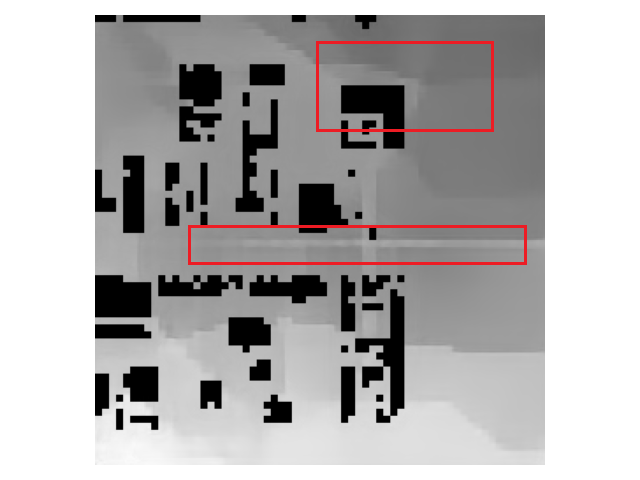}}
        \hspace{-3.18em} 
    \subfigure[RMTransformer]{
        \includegraphics[width=0.27\textwidth]{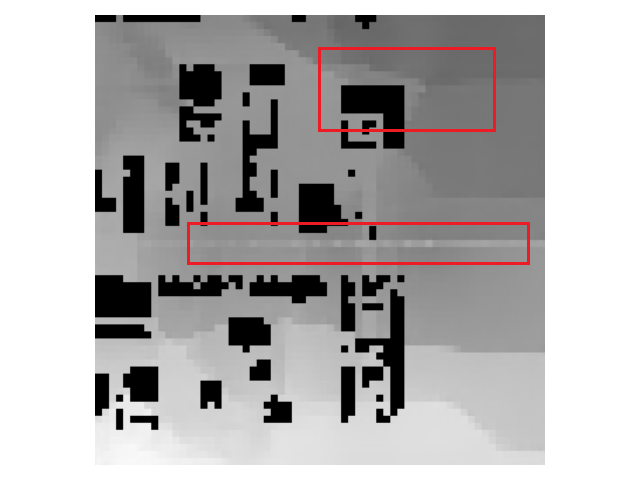}}
        \hspace{-3.18em} 
    \subfigure[Ground truth]{
        \includegraphics[width=0.27\textwidth]{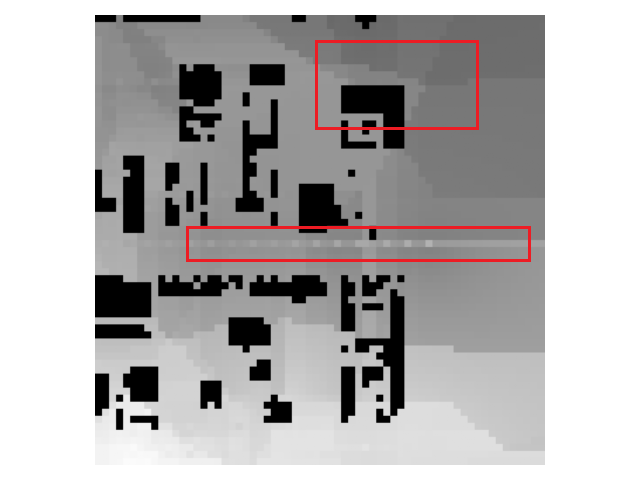}}
    \subfigure[RadioUnet]{
        \includegraphics[width=0.27\textwidth]{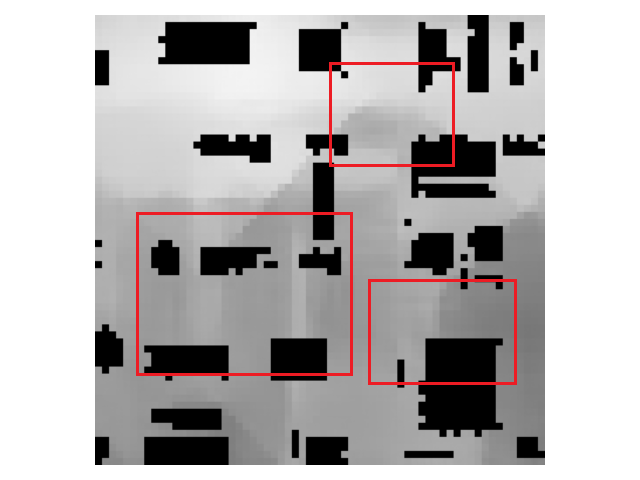}}
        \hspace{-3.18em} 
    \subfigure[PMNet]{
        \includegraphics[width=0.27\textwidth]{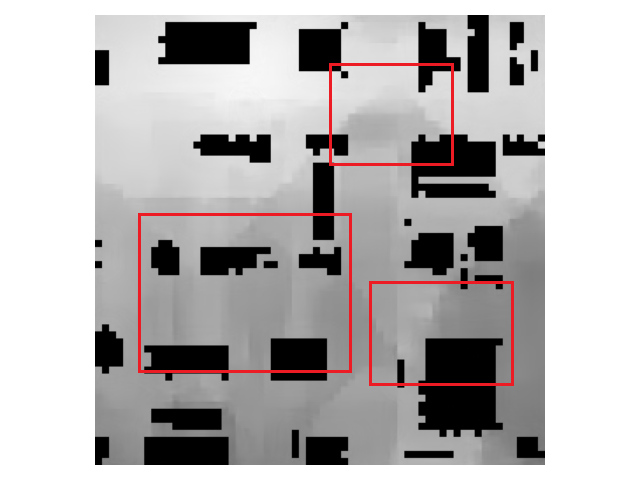}}
        \hspace{-3.18em} 
    \subfigure[RMTransformer]{
        \includegraphics[width=0.27\textwidth]{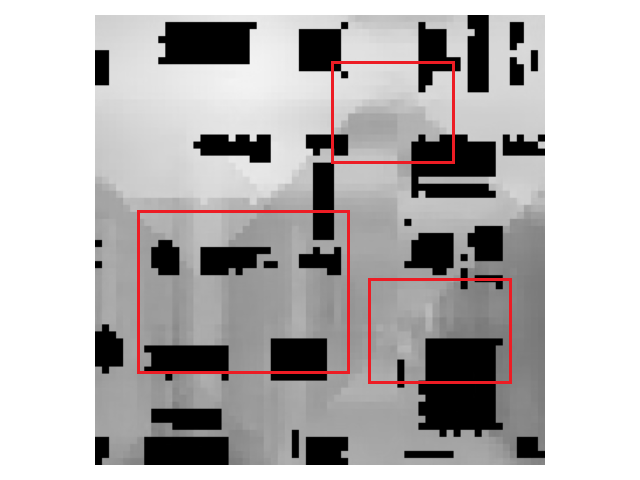}}
        \hspace{-3.18em} 
    \subfigure[Ground truth]{
        \includegraphics[width=0.27\textwidth]{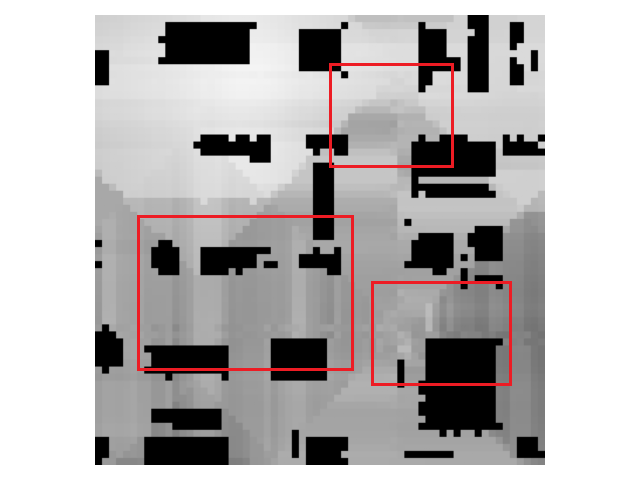}}
    \caption{Visualization of predicted radio maps.}
    \label{fig.visual_usc}
\end{figure*}

\section{Simulations and Discussion}
\subsection{Simulation Setup}
We train and evaluate the models on the USC dataset \cite{lee_scalable_2024}, which is constructed based on the geographical maps in University of Southern California (USC) and ray tracing-based channel emulation \cite{wireless_insite}. The dataset contains a total of $19,016$ samples, which are randomly split into training and testing sets with proportions of $90\%$ and $10\%$, respectively.
Each image has a size of $256 \times 256$, with each pixel corresponding to a grid area of $0.86\,\text{m} \times 0.86\,\text{m}$ in the physical environment.
The transmitted power is set to $0$ dBm, i.e., the received power in dBm and pathloss in dB are identical in value. Received power and pathloss values are normalized between $0$ and $1$, with the minimum value of $-254$ dBm and the maximum value of $0$ dBm to fill a gray-scale image as in \cite{lee_scalable_2024}, while the value $0$ represents the buildings or blockages out of the RoI.
Four ViT blocks and CNN blocks are utilized in the encoder and decoder, respectively, and the specific neural network structure is listed in Table \ref{tab.net}. We train the model for $50$ epochs using the Adam optimizer, where the batch size is $8$, and the learning rate decays from $10^{-4}$ to $10^{-5}$.

\subsection{Evaluation Metrics}
\subsubsection{Root mean square error (RMSE)}
RMSE measures the pixel-level difference between the predicted radio map and the ground truth over the testing samples, as defined in \cite{lee_scalable_2024}.

\subsubsection{Channel prediction error (Ch. Pred. Err.)}
This metric evaluates the RMSE of predicted received powers within the RoI, excluding buildings and blockages, as specified in \cite{lee_scalable_2024}.

\subsubsection{Coverage prediction error (Cov. Pred. Err.)}
This metric assesses whether the predicted received power $\hat{P}_{\mathrm{RX}}(i, j)$ and the ground truth ${P}_{\mathrm{RX}}(i, j)$ at a pixel within the RoI are simultaneously above or below the coverage threshold ${P}_{\text{thres}}$, which is computed as

\begin{equation}
  \text{Cov. Pred. Err.} = \frac{\sum_{i=1}^{W}\sum_{j=1}^{H}E^{P}_{i,j}}{N_{\text{RoI}}},
\end{equation}
where $N_{\text{RoI}}$ is the number of pixels in the RoI, and the binary indicator $E^{P}_{i,j}=0$ if $\hat{P}_{\mathrm{RX}}(i, j), {P}_{\mathrm{RX}}(i, j) \geq {P}_{\text{thres}}$ or $\hat{P}_{\mathrm{RX}}(i, j), {P}_{\mathrm{RX}}(i, j) < {P}_{\text{thres}}$. Otherwise, $E^{P}_{i,j}=1$.
Correct predictions on whether a specific UE is covered, i.e., whether the received power or user rate is desirable, could help us to plan the network deployment or proactively allocate the wireless resources with reduced cost.

\subsection{Numerical Results} 
We train and evaluate the proposed RMTransformer, as well as existing SOTA schemes, i.e., RadioUnet \cite{levie_radiounet_2021} and PMNet \cite{lee_scalable_2024}, using their originally reported network settings. A comparison of the derived RMSEs, channel prediction errors, and coverage prediction errors is listed in Table \ref{tab.nmse}, with the coverage prediction threshold set to 0.8. Benefiting from the hybrid transformer-CNN structure and the desirable feature representation capability of transformers, the proposed RMTransformer significantly outperforms existing SOTA schemes with RMSE reductions of $62.46\%$ over RadioUnet and $31.66\%$ over PMNet.

Fig. \ref{fig.visual_usc} visualizes the predicted radio maps produced by different schemes, shown as gray-scale images. In these images, lighter shades indicate higher received power, while black regions correspond to non-RoI buildings and blockages. Due to the complicated physical environment in wireless propagation related to various factors, e.g., fading, scattering, diffraction and reflection, etc., the radio map patterns could be rather complex to model. Compared to RadioUnet and PMNet, the proposed RMTransformer can effectively recover some complicated details in the pathloss patterns, particularly in areas marked by red rectangles in Fig. \ref{fig.visual_usc}. This is most evident at the boundaries between regions with differing propagation characteristics, enabling more precise segmentation of signal strength levels. Therefore, the proposed RMTransformer not only improves pixel-level prediction accuracy, but also helps to determine the coverage and user signal strength status.

\section{Conclusion}
In this paper, we proposed the RMTransformer, a hybrid transformer-CNN architecture designed for constructing radio maps from geographical inputs. Leveraging the powerful capability in feature representation, we designed a multi-scale transformer-based encoder to effectively extract the features from the geographical maps, while the pixel-level radio map reconstruction is then completed via the CNN-based decoder. Simulation results revealed significant gains over existing SOTA schemes in prediction errors, demonstrating its effectiveness in accurately understanding the wireless propagation features. Future works may include DT-enabled network management and optimization, which takes the predicted radio map as a virtual replica for pre-validation and safe-exploration.

\bibliographystyle{IEEEtran}
\bibliography{./pmap.bib}

\end{document}